\documentclass[10pt,twocolumn,twoside]{article}
\usepackage{amssymb}
\usepackage{amsmath}
\usepackage[dvips]{graphicx}
\usepackage{graphicx}
\usepackage[latin1]{inputenc}

\begin{document}

\date{}
\title{Work and energy in rotating systems}
\author{Diego A. Manjarrés\thanks{%
damanjarrnsg@unal.edu.co}, William J. Herrera\thanks{%
jherreraw@unal.edu.co}, Rodolfo A. Díaz\thanks{%
radiazs@unal.edu.co}. \\
Universidad Nacional de Colombia,\\
Departamento de Física. Bogotá, Colombia.}
\maketitle

\begin{abstract}
{\footnotesize Literature analyzes the way in which Newton's second law can
be used when non-inertial rotating systems are used. However, the treatment
of the work and energy theorem in rotating systems is not considered in
textbooks. In this paper, we show that the work and energy theorem can still
be applied to a closed system of particles in a rotating system, as long as
the work of fictitious forces is properly included in the formalism. The
coriolis force does not contribute to the work coming from fictitious
forces. It worths remarking that real forces that do not do work in an
inertial reference frame can do work in the rotating reference frame and
viceversa. The combined effects of acceleration of the origin and rotation
of the non-inertial system are also studied.}


\end{abstract}


\section{Introduction.}

Most of our laboratory reference frames are non-inertial, for instance, the
dynamics of air or water clusters is usually studied from a reference frame
attached to the earth, this dynamics is strongly determined by the presence
of the coriolis term\cite{Kleppner,Goldstein}. Meteorological and
oceanographic phenomena are also influenced by fictitious forces generated
by considering the earth as a rotating system. Further, the work and energy
formalism could have some advantages in rotating systems, similar to the
case of inertial reference frames. Consequently, it is important to study
the transformation properties of work and energy quantities from an inertial
reference frame to a frame in relative rotation with respect to the inertial
one.

The transformation properties of work and energy between reference frames in
relative translation have been studied \cite{AJPKap, EJPCamarca, American}.
On the other hand, the transformation of forces between an inertial and a
rotating reference frames is quite well studied in most of the literature
(e.g. Ref. \cite{Kleppner,Goldstein}) . Nevertheless, the transformation
properties of work and energy quantities between reference frames in
relative rotation have not been considered. The latter is important since
fictitious forces arising from rotating systems could have potential
energies associated, and real forces that do not do work in a given inertial
frame can do work in the rotating frame.\newline

In Ref. \cite{American}, it was shown that the work and energy theorem is
covariant between inertial reference frames, and that the theorem still
holds in non-inertial translational frames if the works done by fictitious
forces are included appropriately. Further, it was shown that fictitious
forces can do work, and that even in transformations between inertial frames
forces that do not do work in an inertial frame can do work in another
inertial frame obtaining a non-trivial potential energy (see also Ref. \cite%
{Wolfram}). In this work we extend the study of the work and energy
formalism for non-inertial systems that combine accelerated translation with
rotation with respect to an inertial frame. The main results are illustrated
with a pedagogical example, where we solve the problem in both an inertial
and a non-inertial frames, and we show a force that does do work in the
inertial frame, but does not do work in the non-inertial frame. The latter
fact simplifies the problem considerably when treated in the non-inertial
frame.


\section{Formulation of the problem.}

The system under study consists of $N$ interacting particles. The system is
closed, i.e. there is not any flux of particles from or toward the system
and the number of particles is conserved so that not processes of creation
or destruction of particles are considered. However, the system could be
under the influence of time dependent (or time independent) external forces.
Our description is non-relativistic such that time and the mass of the
particles are independent of the reference frame.\newline

\begin{figure}[tbh]
\centering\includegraphics[scale=0.8]{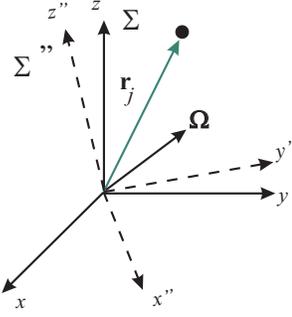} \vspace{0.4cm}
\caption{\emph{Position of a particle from the point of view of (1) the
inertial frame }$\Sigma $\emph{, (2) the non-inertial frame }$\Sigma
^{\prime \prime }$\emph{\ that rotates with angular velocity }$\Omega $\emph{%
\ with respect to }$\Sigma $\emph{. The vector}$\ \Omega $\emph{\ is defined
with a right-hand convention.}}
\label{sistemas}
\end{figure}

Let us define an inertial system $\Sigma $, and a non-inertial rotating
system $\Sigma ^{\prime \prime }$ with constant angular velocity\ $\mathbf{%
\Omega }$ with respect to $\Sigma $ and a common origin between them. With
respect to $\Sigma $, a given $j-th\ $particle has a position, velocity and
acceleration $\mathbf{r}_{j},\mathbf{v}_{j},\mathbf{a}_{j}$. These variables
measured by $\Sigma ^{\prime \prime }$ are denoted by $\mathbf{r}%
_{j}^{\prime \prime },\mathbf{v}_{j}^{\prime \prime },\mathbf{a}_{j}^{\prime
\prime }$. Since $\Sigma $ and $\Sigma ^{\prime \prime }$ have a common
origin, we see that 
\begin{equation*}
\mathbf{r}_{j}^{\prime \prime }=\mathbf{r}_{j}.
\end{equation*}%
Figure \ref{sistemas}, illustrates these statements. We shall analyze the
way in which work and energy transform when we pass from the inertial system 
$\Sigma $, to the non-inertial system $\Sigma ^{\prime \prime }$. The
relationship between velocities and accelerations in $\Sigma $ and $\Sigma
^{\prime \prime }$ is well known from the literature \cite{Kleppner}

\begin{eqnarray}
\mathbf{v}_{j}^{\prime\prime} &=&\mathbf{v}_{j}-\mathbf{\Omega}\times\mathbf{%
r}_{j},  \label{vpjsimple} \\
\mathbf{a}_{j}^{\prime\prime} &=&\mathbf{a}_{j}-2\mathbf{\Omega}\times%
\mathbf{v}_{j}^{\prime\prime}-\mathbf{\Omega}\times\left(\mathbf{\Omega}%
\times\mathbf{r}_{j}\right),  \label{apjsimple}
\end{eqnarray}

multiplying Eq. (\ref{apjsimple}) by the mass $m_{j}$ we find

\begin{eqnarray}
m_{j}\frac{d\mathbf{v}_{j}^{\prime \prime }}{dt} &=&\mathbf{F}_{j}+\mathbf{F}%
_{j,fict},  \label{mjvppjsimple} \\
\mathbf{F}_{j,fict} &=&\mathbf{F}_{j,cor}+\mathbf{F}_{j,cent},
\label{mjvppj1simple} \\
\mathbf{F}_{j,cor} &\equiv &-2m_{j}\mathbf{\Omega }\times \mathbf{v}%
_{j}^{\prime \prime },  \label{coriolissimple} \\
\mathbf{F}_{j,cent} &\equiv &-m_{j}\mathbf{\Omega }\times \left( \mathbf{%
\Omega }\times \mathbf{r}_{j}\right) ,  \label{mjvppj2simple}
\end{eqnarray}%
where $\mathbf{F}_{j}$ represents the total real force on the $j-th$
particle (summation of internal and external forces on $j$), $\mathbf{F}%
_{cor}$ and $\mathbf{F}_{cent}$ are the well-known coriolis and centrifugal
forces. Taking the dot product on both sides of Eq. (\ref{mjvppjsimple}) by $%
\mathbf{v}_{j}^{\prime \prime }dt$ on the left and by $d\mathbf{r}%
_{j}^{\prime \prime }$ on the right, and summing over all particles of the
system we have

\begin{eqnarray}
\sum_{j}d\left(\frac{1}{2}m_{j}\mathbf{v}_{j}^{\prime\prime2}\right)&=&%
\sum_{j} \left(\mathbf{F}_{j}+\mathbf{F}_{fict}\right)\cdot d\mathbf{r}%
_{j}^{\prime\prime},  \notag \\
dK^{\prime\prime}&=&dW^{\prime\prime}.  \label{WErotsimple}
\end{eqnarray}

Where $dK^{\prime \prime }$ and $dW^{\prime \prime }$ are the differentials
of kinetic energy and work when an infinitesimal path $d\mathbf{r}%
_{j}^{\prime \prime }$ is taken for each particle. This equation shows the
covariance of the fundamental theorem of work and energy between $\Sigma $
and $\Sigma ^{\prime \prime }$. In relating work and energy observables
between $\Sigma $ and $\Sigma ^{\prime \prime }$ it is important to write $%
dK_{j}^{\prime \prime }$ and $dW_{j}^{\prime \prime }$ (differential of
kinetic energy and work associated with the $j-$th particle in the system $%
\Sigma ^{\prime \prime }$) in terms of quantities measured by $\Sigma $. For 
$dK_{j}^{\prime \prime }$, we use Eq. (\ref{vpjsimple}) to get

\begin{eqnarray}
dK_{j}^{\prime \prime } &=&m_{j}\mathbf{v}_{j}^{\prime \prime }\cdot d%
\mathbf{v}_{j}^{\prime \prime }  \notag \\
&=&m_{j}\left\{ \mathbf{v}_{j}-\mathbf{\Omega }\times \mathbf{r}_{j}\right\}
\cdot \left\{ d\mathbf{v}_{j}-\mathbf{\Omega }\times d\mathbf{r}_{j}\right\}
\notag \\
dK_{j}^{\prime \prime } &=&dK_{j}+dZ_{j},  \label{dKppdzsimple}
\end{eqnarray}%
with%
\begin{equation*}
dZ_{j}\equiv -(\mathbf{\Omega }\times \mathbf{r}_{j})\cdot d\mathbf{P}%
_{j}-m_{j}[\mathbf{\Omega }\times (\mathbf{\Omega }\times \mathbf{r}%
_{j})]\cdot d\mathbf{r}_{j},
\end{equation*}%
where $d\mathbf{P}_{j}$ denotes the differential of linear momentum
associated with the $j-$th particle measured by $\Sigma $. The coriolis
force given by Eq. (\ref{coriolissimple}) does not do work with respect to $%
\Sigma ^{\prime \prime }$.\ To obtain $dW_{j}^{\prime \prime }$ in terms of
variables measured by $\Sigma $, we use Eqs. (\ref{vpjsimple}, \ref%
{mjvppj1simple}-\ref{mjvppj2simple})

\begin{eqnarray}
\mathbf{F}_{j}^{\prime \prime } &=&\mathbf{F}_{j}-m_{j}[2\mathbf{\Omega }%
\times \mathbf{v}_{j}^{\prime \prime }+\mathbf{\Omega }\times (\mathbf{%
\Omega }\times \mathbf{r}_{j})], \\
d\mathbf{r}_{j}^{\prime \prime } &=&\mathbf{v}_{j}^{\prime \prime }dt=d%
\mathbf{r}_{j}-(\mathbf{\Omega }\times \mathbf{r}_{j})dt,  \notag \\
dW_{j}^{\prime \prime } &=&\mathbf{F}_{j}^{\prime \prime }\cdot d\mathbf{r}%
_{j}^{\prime \prime }=\left( \mathbf{F}_{j}+\mathbf{F}_{j,fict}\right) \cdot
d\mathbf{r}_{j}^{\prime \prime }  \notag \\
&=&\{\mathbf{F}_{j}-m_{j}\mathbf{\Omega }\times (\mathbf{\Omega }\times 
\mathbf{r}_{j})\}\cdot \{d\mathbf{r}_{j}-\mathbf{\Omega }\times \mathbf{r}%
_{j}~dt\},  \notag \\
dW_{j}^{\prime \prime } &=&dW_{j}+dZ_{j},  \label{dWppdzsimple}
\end{eqnarray}%
so the covariance of the fundamental theorem of work and energy is expressed
by Eq. (\ref{WErotsimple}), or equivalently by (\ref{dKppdzsimple}, \ref%
{dWppdzsimple}).

For pedagogical reasons the covariance of the work and energy theorem for
the pure rotation case is realized first, for the reader to assimilate the
formalism in a simple way. The additional subtleties that involve the
combination of translation and rotation are introduced in appendix \ref%
{apendicegeneral}, in which we consider a non-inertial system $\Sigma
^{\prime \prime }$, that possesses a relative rotation with time-dependent
angular velocity and translation with respect to $\Sigma $.

\section{Fictitious work}

We shall consider the general case in which $\Sigma ^{\prime \prime }$
rotates and translates with respect to $\Sigma $ (see appendix \ref%
{apendicegeneral}). The work observed by $\Sigma ^{\prime \prime }$ can be
separated in the work coming from real forces and those coming from
fictitious forces

\begin{eqnarray}
dW^{\prime\prime}&=&dW_{real}+dW_{fict},  \label{separaciontrabajos2} \\
dW_{fict}&=&-\sum_{j}m_{j}[\boldsymbol{\Omega}\times(\boldsymbol{\Omega}%
\times\mathbf{r}_{j}^{\prime\prime})+\boldsymbol{\dot{\Omega}}\times\mathbf{r%
}_{j}^{\prime\prime}  \notag \\
&&+\mathbf{A}(t) ]\cdot d\mathbf{r}_{j}^{\prime\prime},
\label{trabajoficticio2} \\
dW_{real}&=&\sum_{j}[\mathbf{F}_{j}^{\prime\prime}+m_{j}\{\mathbf{\Omega}%
\times(\mathbf{\Omega}\times\mathbf{r}_{j}^{\prime\prime})+\dot{\mathbf{%
\Omega}}\times\mathbf{r}_{j}^{\prime\prime}  \notag \\
&&+\mathbf{A}(t)\}]\cdot d\mathbf{r}_{j}^{\prime\prime}.
\label{trabajoreal22}
\end{eqnarray}

Eqs. (\ref{trabajoficticio2}, \ref{trabajoreal22}) are written is terms of
observables measured by $\Sigma ^{\prime \prime }$\ (except for $\mathbf{%
\Omega }$ and $\mathbf{\dot{\Omega}}$ which are measured with respect to $%
\Sigma $). It is because in most of the problems involving non-inertial
systems, experiments are done on the non-inertial system and measured with
respect to it. In particular, real forces that do not do work in $\Sigma $
can do work in $\Sigma ^{\prime \prime }$, or real forces that do work in $%
\Sigma $ could do no work in $\Sigma ^{\prime \prime }$.\newline

For the particular case $\boldsymbol{\Omega }=0$ it can be proved that if $%
\Sigma^{\prime\prime}$ is attached to the center of mass of the system of
particles (CM), the total fictitious work is null \cite{American},

\begin{equation}
dW_{fict}^{CM}=-M\mathbf{A}_{C}(t)\cdot d\left( \frac{\sum_{j}m_{j}\mathbf{r}%
_{j}^{\prime \prime }}{M}\right) =0,  \label{WfictCM}
\end{equation}%
where $\mathbf{A}_{C}$ is the acceleration of the CM. In the most general
case with $\boldsymbol{\Omega }\neq 0$, the total fictitious work measured
by $\Sigma ^{\prime \prime }$ can be different from zero even if its origin
is attached to the CM.\newline

These results are in close analogy with the case of torques analyzed from
systems attached to the CM. For non-rotating systems (with respect to $%
\Sigma $) attached to the CM, the total fictitious torque and work are null 
\cite{American,Mexicana}. On the other hand, for rotating systems attached
to the CM both the total fictitious torque and the total fictitious work are
in general non-vanishing \cite{Mexicana}. It worths saying that even in
non-rotating systems, the fictitious torque and work on each particle are in
general different from zero.\newline

In addition, it is also possible to find non-inertial systems for which
fictitious work is null and fictitious torque is non-null or vice versa.
However, these features depend no only on the reference frame, but also on
the system of particles involved.


\section{Pedagogical example.\label{sec:example}}

\begin{figure}[tbh]
\centering
\includegraphics[scale=0.8]{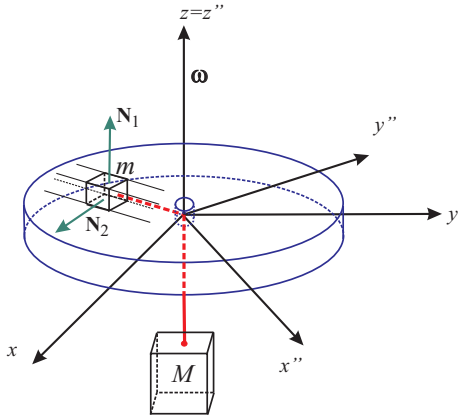} \vspace{0.4cm}
\caption{\emph{A block of mass }$m$\emph{\ slides on a table along a groove,
which rotates with constant angular velocity }$\boldsymbol{\protect\omega }$%
\emph{. The block is attached to other block of mass }$M$\emph{. The Normal
forces }$N_{1}$\emph{\ and }$N_{2}$\emph{\ on the block, are illustrated.}}
\label{ejemplo}
\end{figure}

Let us consider a block of mass $m$ sliding without friction on a horizontal
table which rotates with constant angular velocity $\boldsymbol{\omega }$.
The block is constrained to move on a groove that is radially directed. The
block is attached through a rope of negligible mass to other block of mass $%
M $, which hangs from the table through its center (see Fig. \ref{ejemplo}).
Our physical system of interest are the two blocks, and their description
will be made from the point of view of an inertial reference frame $\Sigma $
whose origin is located at the center of the table, and other non-inertial
system $\Sigma ^{\prime \prime }$ fixed to the table, with a common origin
with $\Sigma $, and with constant angular velocity $\boldsymbol{\omega }$.
The dimensions of the two blocks are neglected, so that we regard the two
blocks as point-like particles.\newline

Our aim is to apply the fundamental work-energy theorem, therefore the work
done by the resultant force due to all applied forces on the system is
calculated. We take in mind that the block of mass $m$, is constrained to
move on a groove. $\mathbf{N}_{2}\ $is a normal force from the wall of the
groove on the mass $m$, which corresponds to the reaction of the table over
the block due to rotation. This force is tangential, and it is different
from the Normal force $\mathbf{N}_{1}$ which is the force on $m$ due to the
vertical contact with the table. It is clear that $\mathbf{N}_{1}$ does not
do work with respect to $\Sigma $ or $\Sigma ^{\prime \prime }$.

We remark the fact that from the point of view of $\Sigma $, the normal
force $\mathbf{N}_{2}$ does work, since there are few examples in the
literature in which normal forces do work \cite{American}. For an observer
in $\Sigma $, the work done by $\mathbf{N}_{2}$ is given by

\begin{equation*}
W_{N}=\int_{\theta _{0}}^{\theta _{f}}N_{2}rd\theta .
\end{equation*}%
from Newton's second law the normal force is given by

\begin{equation*}
\mathbf{N}_{2}=m(r\ddot{\theta}+2\dot{r}\dot{\theta})\mathbf{u}_{\theta
}=2m\omega \dot{r}\mathbf{u}_{\theta }\ ,
\end{equation*}%
so that,

\begin{eqnarray*}
W_{N} &=&m\omega \int_{\theta _{0}}^{\theta _{f}}2r\frac{dr}{dt}d\theta
=m\omega \int_{r_{0}}^{r_{f}}2r~dr\frac{d\theta }{dt} \\
W_{N} &=&m\omega ^{2}\int_{r_{0}}^{r_{f}}2r~dr
\end{eqnarray*}%
hence, the work done by $\mathbf{N}_{2}$ and the total external work (with
respect to $\Sigma $)\ are given by

\begin{equation*}
W_{N}=m\omega ^{2}(r_{f}^{2}-r_{0}^{2})\ \ ,\ \ W=m\omega ^{2}({r_{f}^{2}}-{%
r_{0}^{2}})-Mg(z_{f}-z_{0}).
\end{equation*}

From the point of view of $\Sigma ^{\prime \prime }$, the centrifugal force
on the block of mass $M$ is zero, since $\boldsymbol{\Omega \ }$is\ parallel
to the position of $M$. Furthermore, the coriolis force does not do work.
Therefore, the total fictitious force on $M\ $does not do work. In $\Sigma
^{\prime \prime }$ the displacement of the block of mass $m$ is radial so
that we have%
\begin{equation*}
\boldsymbol{\omega }=\omega ~\mathbf{u}_{z}\ \ ,\ \ \mathbf{r}^{\prime
\prime }=\mathbf{r}=r\mathbf{u}_{r}\ ,\ \ d\mathbf{r}^{\prime \prime
}=dr^{\prime \prime }\mathbf{u}_{r}=dr\ \mathbf{u}_{r}
\end{equation*}%
hence, the work done by $\mathbf{N}_{2}$ is zero in $\Sigma ^{\prime \prime }
$. Therefore, $\mathbf{N}_{2}$ does work in $\Sigma $ but does not do work
in $\Sigma ^{\prime \prime }$. On the other hand, the work done by the
fictitious forces in $\Sigma ^{\prime \prime }\ $is

\begin{equation}
dW^{\prime \prime }=-m\boldsymbol{\omega }\times \left( \boldsymbol{\omega }%
\times \mathbf{r}^{\prime \prime }\right) \cdot d\mathbf{r}^{\prime \prime
}=m\omega ^{2}r^{\prime \prime }dr^{\prime \prime }.
\end{equation}%
The total fictitious work seen by $\Sigma ^{\prime \prime }$ is

\begin{equation}
W^{\prime \prime }=\frac{m\omega ^{2}}{2}({r_{f}^{\prime \prime }}^{2}-{%
r_{0}^{\prime \prime }}^{2})=\frac{m\omega ^{2}}{2}({r_{f}^{2}}-{r_{0}^{2}}).
\end{equation}

The work-energy theorem applied on both $\Sigma $ and $\Sigma ^{\prime
\prime }$, yields

\begin{eqnarray}
W &=&m\omega ^{2}({r_{f}^{2}}-{r_{0}^{2}})-Mg(z_{f}-z_{0})  \notag \\
&=&\frac{m\omega ^{2}}{2}({r_{f}^{2}}-{r_{0}^{2}})+\frac{\left( m+M\right) }{%
2}({v_{r,f}^{2}}-{v_{r,0}^{2}}),  \label{trabajoinercialej} \\
W^{\prime \prime } &=&\frac{m\omega ^{2}}{2}({r_{f}^{2}}-{r_{0}^{2}}%
)-Mg(z_{f}-z_{0})  \notag \\
&=&\frac{\left( m+M\right) }{2}({v_{r,f}^{2}}-{v_{r,0}^{2}}).
\label{trabajorotanteej}
\end{eqnarray}%
We have taken into account that $\mathbf{v}_{M}^{2}=\mathbf{v}_{r}^{2}$
where $\mathbf{v}_{M},~\mathbf{v}_{r}$ are the velocity of $M$ and the
radial velocity of $m$ respectively. The change of kinetic energy of the
system has been separated in tangential and radial parts. For convenience,
we define the radial kinetic energy of the system as the sum of radial
kinetic energy of $m\ $plus the kinetic energy of $M$. We can see from (\ref%
{trabajoinercialej}, \ref{trabajorotanteej}), that the change in radial
kinetic energy is the same in both reference frames. As a consequence, it
could be useful to define an effective work in $\Sigma $, so that the
transversal component of the kinetic energy is included in the work, and
therefore an effective work-energy theorem can be established.\newline

\begin{eqnarray}
W &=&\frac{m\omega ^{2}}{2}({r_{f}^{2}}-{r_{0}^{2}})+W_{ef}=\Delta K_{\theta
}+W_{ef}  \notag \\
W_{ef} &=&\frac{m\omega ^{2}}{2}({r_{f}^{2}}-{r_{0}^{2}})-Mg(z_{f}-z_{0}) 
\notag \\
&=&\frac{\left( m+M\right) }{2}({v_{r,f}^{2}}-{v_{r,0}^{2}})
\label{trabajoeffej}
\end{eqnarray}%
where $\Delta K_{\theta }$ denotes the change in transversal kinetic energy.
The effective work $W_{ef}$ defined in $\Sigma \ $in this way, is equal to
the work $W^{\prime \prime }$ seen by $\Sigma ^{\prime \prime }$ and both
are equal to the change in radial kinetic energy seen by either frame. From $%
W_{ef}$ we can define a effective potential as it is done in the problem of
a central force \cite{Kleppner},

\begin{equation*}
V_{ef}=-\frac{m\omega ^{2}}{2}{r}^{2}+Mgz
\end{equation*}%
Equations (\ref{trabajorotanteej}, \ref{trabajoeffej}) can be rewritten by
taking into account that $z_{f}-z_{0}=r_{f}-r_{0}$

\begin{eqnarray}
W_{ef} &=&W^{\prime \prime }=m\omega _{c}^{2}\left[ \left( \frac{\omega }{%
\omega _{c}}\right) ^{2}\overline{r}-r_{0}\right] (z_{f}-z_{0}),
\label{trabajointer} \\
\omega _{c} &\equiv &\left( \frac{Mg}{mr_{0}}\right) ^{1/2}\ \ ,\ \ 
\overline{r}\equiv \frac{r_{f}+r_{0}}{2}.  \notag
\end{eqnarray}

From Eq. (\ref{trabajointer}) we can give a physical interpretation of the
problem where $\omega _{c}$ is the critical frequency that determines the
sense of motion\footnote{%
The solution from the dynamical point of view to obtain the critical
frecuency is obtained in appendix \ref{ap:dynamics}.}. For the particular
case in which the initial radial velocity of the block $m$ vanishes ($%
v_{r,0}=0$), there are three possible situations: i) For $\omega <\omega
_{c} $, the block $m$ moves toward the center of the table and the block $M$
descends, hence we get$\ \overline{r}<r_{0}$ and $z_{f}-z_{0}<0$, so that a
positive work $W^{\prime \prime }\ $is done on the system; ii) For $\omega
>\omega _{c}$, the block $m$ moves away from the center of the table and the
block $M$ ascends, therefore we$\ $have$\ \overline{r}>r_{0}$ and $%
z_{f}-z_{0}>0$, and a positive work $W^{\prime \prime }\ $is done on the
system; iii) If $\omega =\omega _{c}$, the block $M$ remains at rest and the
radial velocity of $m$ vanishes, then$\ \overline{r}=r_{0}$ and $%
z_{f}-z_{0}=0$, \ and the work done is null. We suggest for the reader to
interpret the equation (\ref{trabajointer}) for the case in which the block
of mass $m$ possesses an initial radial velocity different from zero.\newline


\section{Conclusions.}

We have shown the covariance of the work and energy theorem for a
non-uniform rotational frame, where the effects of acceleration of the
origin and rotation of the non-inertial system are included. This covariance
is complied when the work done for the fictitious forces is properly
included. The coriolis force does not contribute to the work coming from
fictitious forces. We generalize the properties when the reference frame is
attached to the center of mass, where the fictitious work on each particle
is in general different from zero but the total fictitious work is zero for
non-rotating non-inertial systems.

The fact that the work done for a force depends on the reference frame, is
illustrated by means of an example. We solve the problem in an inertial
frame and in a rotating non-inertial frame. By defining an effective work in
the inertial frame we can equate such effective work with the one measured
by the non-inertial rotational frame.

\textbf{Acknowledgments}: This work was supported by División de
Investigaciones de la Universidad Nacional de Colombia (DIB), sede Bogotá.


\appendix

\section{General case.}

\label{apendicegeneral}

\begin{figure}[tbh]
\centering\includegraphics[scale=0.8]{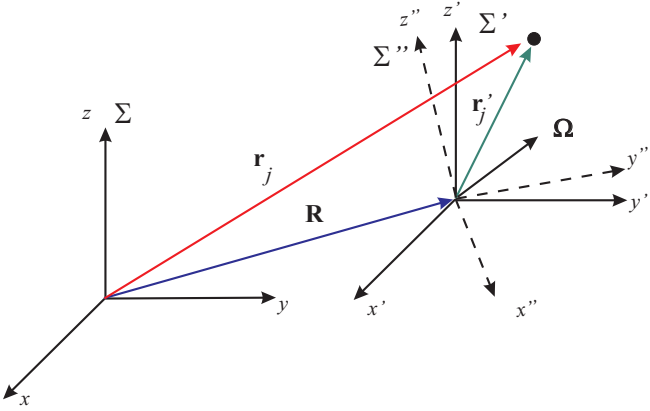}\vspace{0.4cm}
\caption{\emph{Position of a particle from the point of view of three
reference frames (1) the inertial frame }$\Sigma $\emph{, (2) the
non-inertial frame }$\Sigma ^{\prime }$\emph{\ that is in pure translation
with respect to }$\Sigma $\emph{, and (3) The system }$\Sigma ^{\prime
\prime }$\emph{\ that rotates with angular velocity }$\Omega $\emph{\ with
respect to }$\Sigma ^{\prime }$\emph{.}}
\label{sistemasap}
\end{figure}

Let us define an inertial system $\Sigma $, a non-inertial translational and
non-rotating system $\Sigma ^{\prime }$ (with respect to $\Sigma $), and a
rotating system $\Sigma ^{\prime \prime }$ with origin common with $\Sigma
^{\prime }$ and with angular velocity $\mathbf{\Omega }$. The position,
velocity and acceleration of the origin of $\Sigma ^{\prime }$ and $\Sigma
^{\prime \prime }$ with respect to $\Sigma $ are called $\mathbf{R}\left(
t\right) ,\mathbf{V}\left( t\right) ,\mathbf{A}\left( t\right) $. With
respect to $\Sigma $, a given $j-th\ $particle has a position, velocity and
acceleration $\mathbf{r}_{j},\mathbf{v}_{j},\mathbf{a}_{j}$. These variables
measured by $\Sigma ^{\prime }$ are denoted by $\mathbf{r}_{j}^{\prime },%
\mathbf{v}_{j}^{\prime },\mathbf{a}_{j}^{\prime }$ and measured by $\Sigma
^{\prime \prime }$ are denoted by $\mathbf{r}_{j}^{\prime \prime },\mathbf{v}%
_{j}^{\prime \prime },\mathbf{a}_{j}^{\prime \prime }$. Since $\Sigma
^{\prime }$ and $\Sigma ^{\prime \prime }$ have a common origin, we see that 
$\mathbf{r}_{j}^{\prime }=\mathbf{r}_{j}^{\prime \prime }$. The axes of $%
\Sigma $ and $\Sigma ^{\prime }$ are parallel each other at all times.
Figure \ref{sistemasap}, illustrates these statements, this figure also
shows that

\begin{eqnarray}
\mathbf{r}_{j}^{\prime\prime}&=&\mathbf{r}_{j}^{\prime}=\mathbf{r}_{j}-%
\mathbf{R}\left(t\right) ,  \label{rpp=rp} \\
\mathbf{v}_{j}^{\prime}&=&\mathbf{v}_{j}-\mathbf{V}\left(t\right),
\label{vpjvj}
\end{eqnarray}

the relationship between velocities and accelerations in $\Sigma^{\prime}$
and $\Sigma^{\prime\prime}$ is well known from the literature

\begin{eqnarray}
\mathbf{v}_{j}^{\prime\prime} &=&\mathbf{v}_{j}^{\prime}-\mathbf{\Omega}%
\times \mathbf{r}_{j}^{\prime},  \label{vpj} \\
\mathbf{a}_{j}^{\prime\prime}&=&\mathbf{a}_{j}^{\prime}-2\mathbf{\Omega}%
\times\mathbf{v}_{j}^{\prime\prime}-\mathbf{\Omega} \times \left(\mathbf{%
\Omega} \times \mathbf{r}_{j}^{\prime}\right)-\dot{\mathbf{\Omega}}\times 
\mathbf{r}_{j}^{\prime},  \label{apj}
\end{eqnarray}

where we have included the term corresponding to the variation in time of $%
\mathbf{\Omega}$. Combining (\ref{vpjvj}, \ref{vpj}) we have

\begin{equation}
\mathbf{v}_{j}^{\prime \prime }=\mathbf{v}_{j}-\mathbf{\Omega }\times 
\mathbf{r}_{j}^{\prime}-\mathbf{V}\left( t\right),  \label{vppj}
\end{equation}

deriving Eq. (\ref{vpjvj}) with respect to time we find

\begin{equation}
\mathbf{a}_{j}^{\prime }=\mathbf{a}_{j}-\mathbf{A}\left( t\right),
\label{apj2}
\end{equation}

substituting Eq. (\ref{apj2}) in Eq. (\ref{apj}), we obtain

\begin{equation}
\mathbf{a}_{j}^{\prime \prime }=\mathbf{a}_{j}-2\mathbf{\Omega} \times 
\mathbf{v}_{j}^{\prime \prime }-\mathbf{\Omega} \times \left(\mathbf{\Omega}
\times \mathbf{r}_{j}^{\prime }\right) -\dot{\mathbf{\Omega}}\times \mathbf{r%
}_{j}^{\prime }-\mathbf{A}\left(t\right),  \label{appj3}
\end{equation}

multiplying Eq. (\ref{appj3}) by the mass $m_{j}$ we find

\begin{eqnarray}
m_{j}\frac{d\mathbf{v}_{j}^{\prime\prime}}{dt}&=&\mathbf{F}_{j}+\mathbf{F}%
_{j,fict},  \label{mjvppj} \\
\mathbf{F}_{j,fict}&=&\mathbf{F}_{j,cor}+\mathbf{F}_{j,cent}+\mathbf{F}%
_{j,azim}+\mathbf{F}_{j,tras},  \label{mjvppj1} \\
\mathbf{F}_{j,cor}&\equiv&-2m_{j}\mathbf{\Omega}\times\mathbf{v}%
_{j}^{\prime\prime},  \label{coriolis} \\
\mathbf{F}_{j,cent}&\equiv&-m_{j}\mathbf{\Omega}\times\left(\mathbf{\Omega}%
\times\mathbf{r}_{j}^{\prime}\right), \\
\mathbf{F}_{j,azim}&\equiv&-m_{j}\dot{\mathbf{\Omega}}\times\mathbf{r}%
_{j}^{\prime}, \\
\mathbf{F}_{j,tras}&\equiv&-m_{j}\mathbf{A}\left(t\right).  \label{mjvppj2}
\end{eqnarray}

In comparison with Eqs. (\ref{mjvppjsimple}-\ref{mjvppj2simple}), two
additional terms appear, $\mathbf{F}_{azim}$ is the fictitious force coming
from the angular acceleration of $\Sigma ^{\prime \prime }$ with respect to $%
\Sigma $, and $\mathbf{F}_{tras}$ is the term coming from the linear
acceleration of the origin of $\Sigma ^{\prime \prime }$ with respect to $%
\Sigma $. Taking the dot product on both sides of Eq. (\ref{mjvppj}) by $%
\mathbf{v}_{j}^{\prime \prime }dt$ on the left and by $d\mathbf{r}%
_{j}^{\prime \prime }$ on the right, and summing over all particles of the
system, we obtain the covariance of the fundamental work energy theorem
expressed by Eq. (\ref{WErotsimple}), in the general case.\newline

The Coriolis force given by equation (\ref{coriolis}) does not do work with
respect to $\Sigma ^{\prime \prime }$. The differentials of kinetic energy
and work, can be written in terms of quantities measured by $\Sigma $, so

\begin{eqnarray*}
dK_{j}^{\prime\prime}=m_{j}\left\{\mathbf{v}_{j}-\mathbf{\Omega}\times\left[%
\mathbf{r}_{j}-\mathbf{R}\left(t\right)\right] -\mathbf{V}%
\left(t\right)\right\}\cdot  \notag \\
\left\{d\mathbf{v}_{j}-\mathbf{\Omega}\times\left[d\mathbf{r}_{j}-\mathbf{V}%
\left(t\right)dt\right]-d\mathbf{\Omega} \times\left[\mathbf{r}_{j}-\mathbf{R%
}\left(t\right)\right]-\mathbf{A} \left(t\right)dt\right\},
\end{eqnarray*}

\begin{eqnarray*}
dW_{j}^{\prime\prime}=\mathbf{F}_{j}^{\prime\prime}\cdot d\mathbf{r}%
_{j}^{\prime\prime}=(\mathbf{F}_{j}+\mathbf{F}_{j,fict})\cdot\mathbf{v}%
_{j}^{\prime\prime}dt, \\
=\{\mathbf{F}_{j}-m_{j}[\mathbf{\Omega}\times[\mathbf{\Omega}\times( \mathbf{%
r}_{j}-\mathbf{R}(t))]+\dot{\mathbf{\Omega}}\times(\mathbf{r}_{j}-\mathbf{R}%
(t)) \\
+\mathbf{A}(t)]\}\cdot \{d\mathbf{r}_{j}-[\mathbf{\Omega}\times(\mathbf{r}%
_{j}-\mathbf{R}(t))+\mathbf{V}(t)]dt\}.
\end{eqnarray*}

The covariance of the fundamental theorem of work and energy can be
expressed as

\begin{eqnarray}
dK_{j}^{\prime\prime} &=&dK_{j}+dZ_{j},  \label{dKppdz} \\
dW_{j}^{\prime\prime} &=&dW_{j}+dZ_{j},  \label{dWppdz}
\end{eqnarray}

\begin{eqnarray*}
&&dZ_{j} \equiv -[\mathbf{\Omega }\times (\mathbf{r}_{j}-\mathbf{R}(t))+%
\mathbf{V}(t)]\cdot d\mathbf{P}_{j} \\
&&-m_{j}\left\{\mathbf{\Omega}\times\lbrack \mathbf{\Omega}\times (\mathbf{r}%
_{j}-\mathbf{R}(t))]+\dot{\mathbf{\Omega }}\times (\mathbf{r}_{j}-\mathbf{R}%
(t))+\mathbf{A}(t)\right\} \\
&&\cdot \left\{d\mathbf{r}_{j}-[\mathbf{\Omega }\times (\mathbf{r}_{j}-%
\mathbf{R}(t))+\mathbf{V}(t)]dt\right\} .
\end{eqnarray*}


\section{Dynamics.\label{ap:dynamics}}

In the example exposed in Sec. \ref{sec:example}, we can determine the
condition for the angular velocity so that if the block $m$ starts with null
radial velocity ($v_{r,0}=0$), its radial velocity remains null at all
times. From the point of view of $\Sigma $, we analyze the dynamics of the
system composed by the two blocks. In polar coordinates the equations of
motion for the block of mass $m$ are

\begin{eqnarray}
-T=m(\ddot{r}-r\omega^{2}),  \label{eqradialm} \\
N_{2}=2m\omega\dot{r}.  \label{eqtangencialm}
\end{eqnarray}

For the block of mass $M$ we have

\begin{eqnarray}
T-Mg=M\ddot{z}.  \label{eqM}
\end{eqnarray}

The motion of the system is constrained by

\begin{equation}
r-z=\text{constant}  \label{ligadura}
\end{equation}

We combine (\ref{eqradialm}, \ref{eqM}), and employ (\ref{ligadura}), which
leads to

\begin{equation}
\ddot{r}-\frac{m\omega ^{2}}{M+m}r=-\frac{Mg}{M+m}  \label{eqmov}
\end{equation}

The solution of Eq. (\ref{eqmov}), is obtained from the solutions of the
homogeneous equation, hence

\begin{eqnarray}
r &=&C+\left( r_{0}-C\right) \cosh {(\kappa t)}+\frac{v_{r,0}}{\kappa }\sinh 
{(\kappa t)}, \\
\kappa &=&\omega \sqrt{\frac{m}{m+M}}\ \ \ \ ,\ \ \ C=\frac{Mg}{m\omega ^{2}}%
.
\end{eqnarray}

For $v_{r,0}=0$, we can see that if $\omega =\omega _{c}=\left( \frac{Mg}{%
mr_{0}}\right) ^{1/2}$, then the block remains at rest.\newline


\end{document}